\documentclass[english,aps,preprint,superscriptaddress]{revtex4-1}
\usepackage{lmodern}
\usepackage{lmodern}
\usepackage[T1]{fontenc}
\usepackage{amsmath}
\usepackage{amssymb}
\usepackage{graphicx}
\usepackage{esint}

\makeatletter

\usepackage{amsthm}\usepackage{latexsym}\usepackage{bm}\usepackage{amsfonts}

\usepackage{babel}

\usepackage{babel}

\usepackage{hyperref}

\makeatother

\usepackage{babel}
\begin{document}

\title{Field theory and structure-preserving geometric particle-in-cell
algorithm for drift wave instability and turbulence}

\author{Jianyuan Xiao}

\affiliation{School of Physical Sciences, University of Science and Technology
of China, Hefei, 230026, China}

\author{Hong Qin}
\email{hongqin@princeton.edu}

\affiliation{Plasma Physics Laboratory, Princeton University, Princeton, NJ 08543,
U.S.A}

\affiliation{School of Physical Sciences, University of Science and Technology
of China, Hefei, 230026, China}
\begin{abstract}
A field theory and the associated structure-preserving geometric Particle-In-Cell
(PIC) algorithm are developed to study low frequency electrostatic
perturbations with fully kinetic ions and adiabatic electrons in magnetized
plasmas. The algorithm is constructed by geometrically discretizing
the field theory using discrete exterior calculus, high-order Whitney
interpolation forms, and non-canonical Hamiltonian splitting method.
The discretization preserves the non-canonical symplectic structure
of the particle-field system, as well as the electromagnetic gauge
symmetry. As a result, the algorithm is charge-conserving and possesses
long-term conservation properties. Because drift wave turbulence and
anomalous transport intrinsically involve multi time-scales, simulation
studies using fully kinetic particle demand algorithms with long-term
accuracy and fidelity. The structure-preserving geometric PIC algorithm
developed adequately servers this purpose. The algorithm has been
implemented in the \textsl{SymPIC} code, tested and benchmarked using
the examples of ion Bernstein waves and drift waves. We apply the
algorithm to study the Ion Temperature Gradient (ITG) instability
and turbulence in a 2D slab geometry. Simulation results show that
at the early stage of the turbulence, the energy diffusion is between
the Bohm scaling and gyro-Bohm scaling. At later time, the observed
diffusion is closer to the gyro-Bohm scaling, and density blobs generated
by the rupture of unstable modes are the prominent structures of the
fully developed ITG turbulence.
\end{abstract}

\keywords{Structure-preserving geometric algorithm, particle-in-cell, drift
wave instability, ion temperature gradient turbulence}

\pacs{52.65.Rr, 52.25.Dg}

\maketitle
\global\long\def\EXP{\times10^}  
\global\long\def\rmd{\mathrm{d}}  
\global\long\def\rmc{\mathrm{c}}  
\global\long\def\diag{\textrm{diag}}  
\global\long\def\xs{ \mathbf{x}_{sp}}  
\global\long\def\bfx{\mathbf{x}}  
\global\long\def\bfd{\mathbf{d}}  
\global\long\def\bfp{\mathbf{p}}  
\global\long\def\bfv{\mathbf{v}}  
\global\long\def\bfA{\mathbf{A}}  
\global\long\def\bfY{\mathbf{Y}}  
\global\long\def\bfB{\mathbf{B}}  
\global\long\def\bfS{\mathbf{S}}  
\global\long\def\bfG{\mathbf{G}}  
\global\long\def\bfE{\mathbf{E}}  
\global\long\def\bfM{\mathbf{M}}  
\global\long\def\bfQ{\mathbf{Q}}  
\global\long\def\bfu{\mathbf{u}}  
\global\long\def\bfe{\mathbf{e}}  
\global\long\def\bfzig{\mathbf{r}_{\textrm{zig2}}}  
\global\long\def\bfxzig{\mathbf{r}_{\textrm{xzig}}}  
\global\long\def\bfzzig{\mathbf{r}_{\textrm{zzig}}}  
\global\long\def\xzig{\mathbf{x}_{\textrm{zig}}}  
\global\long\def\yzig{\mathbf{y}_{\textrm{zig}}}  
\global\long\def\zzig{\mathbf{z}_{\textrm{zig}}}  
\global\long\def\zigspmvar{\left( \bfx_{sp,l-1},\bfx_{sp,l},\tau \right)}  
\global\long\def\zigspvar{\left( \bfx_{sp,l},\bfx_{sp,l+1},\tau \right)}  
\global\long\def\rme{\mathrm{e}}  
\global\long\def\rmi{\mathrm{i}}  
\global\long\def\rmq{\mathrm{q}}  
\global\long\def\ope{\omega_{pe}}  
\global\long\def\oce{\omega_{ce}}  
\global\long\def\FIG#1{Fig.~\ref{#1}}  
\global\long\def\TAB#1{Tab.~\ref{#1}}  
\global\long\def\EQ#1{Eq.~(\ref{#1})}  
\global\long\def\SEC#1{Sec.~\ref{#1}}  
\global\long\def\APP#1{Appendix~\ref{#1}}  
\global\long\def\REF#1{Ref.~\cite{#1}}  
\global\long\def\DDELTAT#1{\textrm{Dt}\left(#1\right)}  
\global\long\def\DDELTATA#1{\textrm{Dt}^*\left(#1\right)}  
\global\long\def\GRADD{ {\nabla_{\mathrm{d}}}}  
\global\long\def\CURLD{ {\mathrm{curl_{d}}}}  
\global\long\def\DIVD{ {\mathrm{div_{d}}}}  
\global\long\def\CURLDP{ {\mathrm{curl_{d}}^{*}}}  
\global\long\def\DIVDP{ {\mathrm{div_{d}}^{*}}}  
\global\long\def\cpt{\captionsetup{justification=raggedright }}  
\global\long\def\act{\mathcal{A}}  
\global\long\def\calL{\mathcal{L}}  
\global\long\def\calJ{\mathcal{J}}  
\global\long\def\DELTAA{\left( \bfA_{J,l}-\bfA_{J,l}' \right)}  
\global\long\def\DELTAAL{\left( \bfA_{J,l-1}-\bfA_{J,l-1}' \right)}  
\global\long\def\ADAGGER{\bfA_{J,l}^\dagger}  
\global\long\def\ADAGGERA#1{\bfA_{J,#1}^{x/2}}  
\global\long\def\EDAGGER#1{\bfE_{J,#1}^{x/2}}  
\global\long\def\BDAGGER#1{\bfB_{J,#1}^{x/2}}  
\global\long\def\DDT{\frac{\partial}{\partial t}}  
\global\long\def\DBYDT{\frac{\rmd}{\rmd t}}  
\global\long\def\DBYANY#1{\frac{\partial }{\partial #1}} 
\newcommand{\WZERO}[1]{W_{\sigma_0 I}\left( #1 \right)} 
\newcommand{\WONE}[1]{W_{\sigma_1 J}\left( #1 \right)} 
\newcommand{\WONEJp}[1]{W_{\sigma_1 J'}\left( #1 \right)} 
\newcommand{\WTWO}[1]{W_{\sigma_2 K}\left( #1 \right)} 
\newcommand{\WTHREE}[1]{W_{\sigma_3 L}\left( #1 \right)} 
\newcommand{\bfJ}{\mathbf{J}} 
\global\long\def\MQQ{M_{00}} 
\global\long\def\MDQDQ{M_{11}} 
\global\long\def\MDQQ{M_{01}}

\section{Introduction}

Drift wave turbulence and associated anomalous transport \cite{Krall1965,Coppi1967,Tang1978,horton1999drift}
are important physical processes in magnetic fusion devices. They
have been intensively studied using the Particle-In-Cell (PIC) methods
\cite{Dawson1983,hockney1988computer,birdsall1991plasma}, which numerically
solve the Vlasov-Maxwell or Vlasov-Poisson equations. The dynamics
of charged particles in a magnetic field described by the Vlasov equation
contains multiple timescales, e.g., the cyclotron frequencies of electrons
and ions, plasma frequency, and the drift wave frequency. When simulating
low frequency phenomena directly using the PIC method, the time-step
must be chosen small enough to resolve the high frequency dynamics
of charged particles. Thus, the total number of time-steps required
is large, often exceeding computer resource available. The low frequency
drift wave instability is such a case, where the ratio between wave
frequency and the electron gyro-frequency is in the order of $10^{-5}$.
To overcome this difficulty, simplified models which eliminate some
of the high-frequency processes while properly describing the slow
ion dynamics are developed. A commonly adopted such kinetic model
is based on adiabatic electron assumption and quasi-neutrality condition.
In this model, for plasmas with one ion species, the ion density $n_{i}$,
electrons density $n_{e}$, and the electrostatic potential $\phi$
are linked as
\begin{eqnarray}
-\frac{q_{i}}{q_{e}}n_{i}=n_{e}=n_{e0}\exp\left(-\frac{q_{e}\phi}{T_{e}}\right)~,\label{EqnABAE}
\end{eqnarray}
where $q_{i}$ and $q_{e}$ are the charges of ions and electrons,
and $T_{e}$ is the electron temperature. The ion dynamics is governed
by Newton's equation with the Lorentz force, 
\begin{eqnarray}
\ddot{\bfx}_{p} & = & \frac{q_{i}}{m_{i}}\left[\bfE\left(\bfx_{p},t\right)+\dot{\bfx}_{p}\times\bfB_{0}\left(\bfx_{p},t\right)\right]~.\label{EqnDPPX}
\end{eqnarray}
where $\bfB_{0}$ is the background magnetic field, $\bfE=-\nabla\phi+\bfE_{0}$
is the electric field, and $\bfE_{0}$ is the background electric
field.

To further decrease computational complexity, gyrokinetic particle
simulation methods have been developed and applied to study low frequency
instabilities and turbulent transport \cite{Lee1983,dimits1993partially,parker1993fully,parker1993gyrokinetic,hu1994generalized,dimits1996scalings,lin1998turbulent,parker1999large}.
In spite of the success of gyrokinetic simulations, it was pointed
out recently that the basic ordering of the gyrokinetic theory \cite{Frieman66,Catto78,frieman1982nonlinear,Dubin83-3524,Hahm88-2670,Brizard89-541,Qin98-1035,Qin00-4433,Qin04-1052,QinFields,qin2007geometric,BurbyThesis,burby2014hamiltonian}
is not always valid in certain parameter regimes for modern magnetic
fusion devices, especially for the H-mode pedestal physics \cite{wan2012z,wan2013global}
and when density perturbations are large \cite{deng2015testing}.
Moreover, due to the requirement of accuracy and numerical stability,
the time-step $\Delta t$ in gyrokinetic simulations are often restricted
to the same order of ion gyro-period \cite{chen2008coarse,chowdhury2016particle}
already. In these situations, the gyrokinetic method has no significant
computational advantage over fully kinetic methods. Recently, a fully
kinetic ion scheme was developed \cite{miecnikowski2018nonlinear,sturdevant2017low,hu2018fully}.
However, even though adiabatic electron model removes the fast electron
dynamics from the system, drift wave instabilities and turbulence
still evolve in a slow timescale, about one thousandth of ion gyro-period.
Simulating these low frequency physics using fully kinetic ions requires
a large number of time-steps, and the long-term conservative properties
of the numerical schemes become crucial. Conventional PIC methods
are based on direct discretization of differential equations, for
which numerical errors in general accumulate coherently during the
iterations, and long-term simulation results are not reliable.

In the present study, we use a very different approach to construct
an explicit high-order structure-preserving geometric PIC algorithm
for simulating low frequency drift wave instabilities and turbulence
in magnetized plasmas. First, a field theory for low frequency electrostatic
dynamics is established with fully kinetic ions and adiabatic electrons.
Then the field theory is geometrically discretized using Discrete
Exterior Calculus (DEC) \cite{hirani2003discrete,Desbrun2005}, Whitney
interpolating forms \cite{whitney1957geometric,Squire4748,squire2012geometric,xiao2015explicit},
and the powerful Hamiltonian splitting method for Vlasov-Maxwell systems
\cite{xiao2015explicit,he2015hamiltonian,he2017explicit}. The resulting
structure-preserving geometric PIC algorithm is able to preserve the
non-canonical symplectic structure associated with the particle-field
system, and numerical results show that the simulation error on the
energy of the system is bounded by a small number for all time-steps.
In addition, the algorithm is gauge independent and thus exactly complies
with the discrete local charge conservation law. The knowledge of
magnetic potentials is not needed, which is convenient in practical.
Furthermore, the algorithm is locally explicit such that it is more
efficient on parallel clusters compared with implicit schemes.

In the last ten years, structure-preserving geometric algorithm has
become an active research topic in plasma physics. Since 1980s, symplectic
integrators for solving Hamiltonian systems have been systematically
studied \cite{Ruth83,Feng85,feng1986difference,feng2010symplectic,Forest90,Channell90,Candy91,Hong02,Tang93,Shang94,Shang99,Sanz-Serna94,marsden1998multisymplectic,Sun2000,marsden2001discrete,hairer2006geometric}.
The idea of geometric integrators is to find a discrete one-step iteration
map that preserves the symplectic 2-form exactly as the analytical
solution of the Hamiltonian system does. According to theoretical
and numerical investigations, numerical errors of symplectic integrators
on invariants of the systems, such as the total energy and momentum,
can be bounded by small numbers for all time-steps \cite{feng1986difference,Sanz-Serna94,Shang99,hairer2006geometric}.
In plasma physics, many fundamental models are canonical or non-canonical
Hamiltonian systems, and corresponding structure-preserving geometric
integrators were recently developed, including those for guiding centers
\cite{qin2008variational,qin2009variational,zhang2014canonicalization,Ellison2015,Burby2017,Kraus2017,Ellison2018,LelandThesis},
charged particles \cite{qin2013boris,he2015volume,zhang2015volume,ellison2015comment,zhang2016explicit,Wang2016,he2017explicit,He16-172,Tu2016,zhou2017explicit,Xiao2018,Shi2019},
Vlasov-Maxwell systems \cite{Squire4748,squire2012geometric,xiao2013variational,evstatiev2013variational,Shadwick14,xiao2015variational,xiao2015explicit,qin2016canonical,he2016hamiltonian,kraus2017gempic,Morrison2017,xiao2017local,xiao2018structure,Xiao2018a},
ideal two-fluid systems \cite{xiao2016explicit}, magnetohydrodynamics
\cite{zhou2014variational,zhou2015formation,zhou2017explicit,ZhouThesis},
Schr\"{o}dinger-Maxwell system \cite{chen2017canonical} and Klein-Gorden-Maxwell
\cite{Shi2016,Shi2018,ShiThesis} system. One of the defining characteristics
of structure-preserving geometric algorithms is that they are all
based on the underpinning field theories and the geometric discretization
thereof. Structure-preserving geometric algorithms have demonstrated
unparalleled long-term stability and conservative properties compared
with conventional non-geometric methods.

The study reported here represents a new development in this research
field. We customarily design a field theory for low frequency electrostatic
perturbations with fully kinetic ions and adiabatic electrons, and
geometrically discrete the field theory to build a structure-preserving
geometric PIC algorithm for simulating drift wave instabilities and
turbulence in magnetized plasmas.

The paper is organized as follows. In Sec.\,\ref{sec:2}, the field
theory for low frequency electrostatic perturbations with fully kinetic
ions and adiabatic electrons is established, which is the starting
point of our study. Section \ref{sec:3} constructs structure-preserving
geometric PIC algorithm by geometrically discretizing the field theory.
The algorithm is tested using the examples of Ion Bernstein Waves
(IBWs) and drift waves in Sec.\,\ref{sec:4}, and then applied to
study the Ion Temperature Gradient (ITG) instability and turbulence
in a 2D slab geometry.

\section{Field theory for low frequency electrostatic perturbations\label{sec:2}}

To build an effective geometric PIC algorithm for drift wave instabilities
and turbulence, a field theory for low frequency electrostatic perturbations
is required. With the assumptions of adiabatic electrons and quasi-neutrality
condition, the key of establishing the field theory is to find an
appropriate action integral whose Euler-Lagrange (EL) equations recover
Eqs.\,(\ref{EqnABAE}) and (\ref{EqnDPPX}). We have found such an
action integral. It is
\begin{eqnarray}
S\left[\bfx_{p},\phi\right] & = & \int\rmd t\left[\sum_{p}\left(\frac{1}{2}m_{i}|\dot{\bfx}_{p}|^{2}+q_{i}\dot{\bfx}_{p}\cdot\bfA_{0}\left(\bfx_{p},t\right)-q_{i}\phi\left(\bfx_{p},t\right)\right)-\right.\nonumber \\
 &  & \left.\int\rmd\bfx n_{e0}T_{e}\exp\left(-\frac{q_{e}\phi}{T_{e}}\right)\right]~,\label{EqnANAACT}
\end{eqnarray}
where $T_{e}$ and $n_{e0}$ are functions of $\bfx$ and $t$, $\bfA_{0}$
is the external magnetic potential which gives $\bfB_{0}=\nabla\times\bfA_{0}$
and $\bfE_{0}=-\dot{\bfA}_{0}$. The system evolves according to the
EL equations, 
\begin{eqnarray}
\frac{\delta S}{\delta\bfx_{p}} & = & 0~,\label{EqnDSdx}\\
\frac{\delta S}{\delta\phi} & = & 0.\label{EqnDSdp}
\end{eqnarray}
It can be easily verified that Eqs.~(\ref{EqnDSdx}) and (\ref{EqnDSdp})
are equivalent to Eqs.~(\ref{EqnDPPX}) and (\ref{EqnABAE}).

If we insert $\phi$ obtained from \EQ{EqnDSdp} to the action integral
\EQ{EqnANAACT}, then the resulting new action integral is 
\begin{eqnarray}
S'\left[\bfx_{p}\right] & = & \int\rmd t\left[\sum_{p}\left(\frac{1}{2}m_{i}|\dot{\bfx}_{p}|^{2}+q_{i}\dot{\bfx}_{p}\cdot\bfA_{0}\left(\bfx_{p},t\right)\right)-\int\rmd x^{3}\left(\rho\phi\left(\rho\right)-T_{e}n_{e0}\rho\right)\right]\thinspace,\label{eq:6}
\end{eqnarray}
where 
\begin{eqnarray}
\rho\left(\bfx\right) & = & \sum_{p}q_{i}\delta\left(\bfx-\bfx_{p}\right)~,\\
\phi\left(\rho\right) & = & -\frac{T_{e}}{q_{e}}\log\left(-\frac{\rho}{n_{e,0}q_{e}}\right)~.
\end{eqnarray}
The action integral $S'$ does not depend on $\phi$, and the corresponding
EL equation, 
\begin{eqnarray}
\frac{\delta S'}{\delta\bfx_{p}} & = & 0~,\label{EqnDSdx1}
\end{eqnarray}
is equivalent to Eqs.\,(\ref{EqnDSdx}) and (\ref{EqnDPPX}).

By design, the field theory is only applicable to low frequency electrostatic
perturbations with adiabatic electrons. However, it captures the fully
kinetic dynamics of ions, and the corresponding geometric algorithm
constructed in the next section possesses long-term accuracy and fidelity,
a necessity for fully kinetic particle simulations guaranteed only
by the structure-preserving nature of the algorithm.

\section{Structure-preserving geometric PIC algorithm\label{sec:3}}

In previous study, we have built explicit geometric algorithms for
the Vlasov-Maxwell system and ideal two-fluid system \cite{xiao2015explicit,xiao2016explicit,xiao2018structure}.
The action integral given by Eq.\,(\ref{EqnANAACT}) or (\ref{eq:6})
is similar to those in previous work. We therefore apply the same
techniques of DEC \cite{hirani2003discrete,Desbrun2005} and high-order
Whitney interpolating forms \cite{xiao2015explicit,xiao2016explicit}
to perform the spatial discretization. The resulting spatially discretized
action integral is 
\begin{eqnarray}
S_{sd}\left[\bfx_{p},\phi_{I}\right]=\int\rmd tL_{sd}\left[\bfx_{p},\phi_{I}\right]~,
\end{eqnarray}
where 
\begin{eqnarray}
L_{sd}\left[\bfx_{p},\phi_{I}\right] & = & \sum_{p}\left[\frac{1}{2}m_{i}\left|\dot{\bfx}_{p}\right|^{2}+q_{i}\dot{\bfx}_{p}\cdot\bfA_{0}\left(\bfx_{p},t\right)-q_{i}\sum_{I}\WZERO{\bfx_{p}}\phi_{I}\right]-\nonumber \\
 &  & \sum_{I}n_{e0,I}T_{e,I}\exp\left(-\frac{q_{e}\phi_{I}}{T_{e,I}}\right)~\label{EqnLSD}
\end{eqnarray}
is the spatially discretized Lagrangian. Here, the subscript $I$
is the grid index, $T_{e,I}$ and $n_{e0,I}$ are electron temperature
and density fields on the grid, $\WZERO\bfx$ is the Whitney interpolating
map for 0-forms (scalar fields) \cite{xiao2015explicit,xiao2016explicit}.
The equations of motion for the discrete system are
\begin{eqnarray}
\frac{\delta S_{sd}}{\delta\bfx_{p}} & = & 0~,\label{EqnDDSdx}\\
\frac{\delta S_{sd}}{\delta\phi_{I}} & = & 0~.\label{EqnDDSdp}
\end{eqnarray}
Equation (\ref{EqnDDSdp}) plays the role of Poisson's equation, which
links the charge density and the electrostatic potential, i.e., 
\begin{eqnarray}
\phi_{I} & = & -\frac{T_{e,I}}{q_{e}}\log\left(-\frac{\rho_{I}}{n_{e0,I}q_{e}}\right)~,\label{EqnCALPHI}\\
\rho_{I} & = & \sum_{p}q_{i}\WZERO{\bfx_{p}}~.
\end{eqnarray}
Equation (\ref{EqnDDSdx}) is Newton's equation with the Lorentz force
for the $p$-th particle, 
\begin{eqnarray}
\ddot{\bfx}_{p} & = & \frac{q_{i}}{m_{i}}\left[\bfE_{0}\left(\bfx_{p},t\right)+\dot{\bfx}_{p}\times\bfB_{0}\left(\bfx_{p},t\right)-\nabla\sum_{I}\WZERO{\bfx_{p}}\phi_{I}\right]~,\label{EqnDDXdtD}
\end{eqnarray}
where 
\begin{eqnarray}
\bfE_{0} & = & -\dot{\bfA}_{0}~,\\
\bfB_{0} & = & \nabla\times\bfA_{0}~.
\end{eqnarray}
Using the property of Whitney interpolating map \cite{xiao2015variational},
\EQ{EqnDDXdtD} can be rewritten as 
\begin{eqnarray}
\ddot{\bfx}_{p} & = & \frac{q_{i}}{m_{i}}\left(\bfE_{0}\left(\bfx_{p},t\right)+\dot{\bfx}_{p}\times\bfB_{0}\left(\bfx_{p},t\right)+\sum_{J}\WONE{\bfx_{p}}\bfE_{J}\right)~,\label{EqnDDXdtDE}
\end{eqnarray}
where 
\begin{eqnarray}
\bfE_{J}=-\sum_{I}\GRADD_{J,I}\phi_{I}~.
\end{eqnarray}
Akin to the relationship between $S\left[\bfx_{p},\phi\right]$ and
$S'\left[\bfx_{p}\right]$, we can obtain a discrete Lagrangian independent
of $\phi_{I}$ by inserting \EQ{EqnCALPHI} into \EQ{EqnLSD},
\begin{eqnarray}
L_{sd}'\left[\bfx_{p}\right] & = & \sum_{p}\left[\frac{1}{2}m_{i}\left|\dot{\bfx}_{p}\right|^{2}+q_{i}\dot{\bfx}_{p}\cdot\bfA_{0}\left(\bfx_{p},t\right)\right]-V~,\label{EqnLSDPM}\\
V & = & \sum_{I}\left[\rho_{I}\phi_{I}\left(\rho_{I}\right)-T_{e,I}n_{e0,I}\rho_{I}\right]~,\label{EqnLSDP}\\
\rho_{I} & = & \sum_{p}q_{i}\WZERO{\bfx_{p}}~.
\end{eqnarray}

From \EQ{EqnLSDPM} we see that the system now involves only particles.
It is not difficult to build symplectic algorithms for $L_{sd}'\left[\bfx_{p}\right]$
using the techniques of variational integrators \cite{marsden2001discrete,hairer2006geometric}.
However, directly applying these techniques will break the electromagnetic
gauge symmetry of the system, which causes charge accumulation and
results in numerical stability. To overcome this shortcoming, explicit
Hamiltonian splitting method \cite{xiao2015explicit,he2015hamiltonian,he2017explicit,zhou2017explicit}
for charged particle dynamics and the Vlasov-Maxwell system have been
developed. To solve for $L_{sd}'\left[\bfx_{p}\right]$, here we adopt
a similar but more general Hamiltonian splitting method \cite{Xiao2018}.

First, we introduce a non-canonical Hamiltonian structure for the
$p$-th charged particle by extending the phase space into 8-dimensional,
\begin{eqnarray}
H_{p}\left(\bfx_{p},\bfv_{p},W_{p},t_{p}\right)=\frac{1}{2}m_{i}\bfv_{p}^{2}-W_{p}\,.
\end{eqnarray}
The associated Poisson bracket is 
\begin{eqnarray}
\left\{ F,G\right\} _{p}=\nabla_{p}F\left[\begin{array}{cccc}
0 & \frac{1}{m_{i}}I & 0 & 0\\
-\frac{1}{m_{i}}I & \frac{q_{i}}{m_{i}^{2}}\hat{B}_{0}\left(\bfx_{p},t_{p}\right) & \left(\frac{q_{i}}{m_{i}}\frac{\partial\bfA_{0}\left(\bfx_{p},t_{p}\right)}{\partial t_{p}}\right)^{T} & 0\\
0 & -\frac{q_{i}}{m_{i}}\frac{\partial\bfA_{0}\left(\bfx_{p},t_{p}\right)}{\partial t_{p}} & 0 & -1\\
0 & 0 & 1 & 0
\end{array}\right]\left(\nabla_{p}G\right)^{T}~,
\end{eqnarray}
where 
\begin{eqnarray}
\nabla_{p}F & = & \left[\frac{\partial F}{\partial\bfx_{p}},\frac{\partial F}{\partial\bfv_{p}},\frac{\partial F}{\partial W_{p}},\frac{\partial F}{\partial t_{p}}\right]~,\\
\hat{B}_{0} & = & \left[\begin{array}{ccc}
0 & B_{0,z} & -B_{0,y}\\
-B_{0,z} & 0 & B_{0,x}\\
B_{0,y} & -B_{0,x} & 0
\end{array}\right]~.
\end{eqnarray}
The Hamiltonian $H$ and Poisson bracket $\left\{ \cdot,\cdot\right\} $
for the extended system are
\begin{eqnarray*}
H & = & \sum_{p}H_{p}+V~,\\
\left\{ F,G\right\}  & = & \sum_{p}\left\{ F,G\right\} _{p}~,
\end{eqnarray*}
where $V$ is defined in Eq.\,(\ref{EqnLSDP}). Hamilton's equation
is
\begin{eqnarray}
\dot{f} & = & \left\{ f,H\right\} ~,\textrm{for }f\in\left\{ \bfx_{p},\bfv_{p},W_{p},t_{p}\right\} ~.
\end{eqnarray}
Newton's equation with the Lorentz force (\ref{EqnDDXdtD}) is equivalent
to $\dot{\bfv}_{p}=\left\{ \bfv_{p},H\right\} $ and $\dot{\bfx}_{p}=\left\{ \bfx_{p},H\right\} $.

Next, we split the Hamiltonian $H$ into 5 parts,
\begin{eqnarray}
H=H_{x}+H_{y}+H_{z}+H_{W}+H_{V}~,
\end{eqnarray}
where $H_{W}=-\sum_{p}W_{p}$, $H_{V}=V$, $H_{x}=\sum_{p}m_{i}v_{p,x}^{2}/2$,
$H_{y}=\sum_{p}m_{i}v_{p,y}^{2}/2$, and $H_{z}=\sum_{p}m_{i}v_{p,z}^{2}/2$.
Each part represents a sub-Hamiltonian system. For example, the equation
of motion generated by $H_{V}$ is 
\begin{eqnarray}
\dot{f} & = & \left\{ f,H_{V}\right\} ,
\end{eqnarray}
i.e., 
\begin{eqnarray}
\left\{ \begin{array}{ccl}
\dot{\bfx}_{p} & = & 0~,\\
\dot{\bfv}_{p} & = & \partial V/\partial\bfx_{p}~,\\
\dot{W}_{p} & = & 0~,\\
\dot{t}_{p} & = & 0~,
\end{array}\right. & \quad\quad\quad & \textrm{for all }p~.
\end{eqnarray}
Its exact solution map $\Theta_{V}\left(\Delta t\right)$ is 
\begin{eqnarray}
\Theta_{V}\left(\Delta t\right):\left\{ \begin{array}{ccl}
\bfx_{p} & \rightarrow & \bfx_{p}~,\\
\bfv_{p} & \rightarrow & \bfv_{p}+\Delta t\partial V/\partial\bfx_{p}~,\\
W_{p} & \rightarrow & W_{p}~,\\
t_{p} & \rightarrow & t_{p}~,
\end{array}\right. & \quad\quad\quad & \textrm{for all }p~.
\end{eqnarray}
We can exactly solve all other sub-systems$H_{W}$, $H_{x}$, $H_{y}$
and $H_{z}$ in a similar way. The exact solution maps are listed
as follows. 
\begin{eqnarray}
\textrm{For all }p & :\\
\Theta_{W}\left(\Delta t\right) & : & \left\{ \begin{array}{ccl}
\bfx_{p} & \rightarrow & \bfx_{p}~,\\
\bfv_{p} & \rightarrow & \bfv_{p}-\frac{q_{i}}{m_{i}}\left(\bfA_{0}\left(\bfx_{p},t_{p}+\Delta t\right)-\bfA_{0}\left(\bfx_{p},t_{p}\right)\right)~,\\
W_{p} & \rightarrow & W_{p}~,\\
t_{p} & \rightarrow & t_{p}+\Delta t~.
\end{array}\right.\\
\Theta_{x}\left(\Delta t\right) & : & \left\{ \begin{array}{ccl}
\bfx_{p} & \rightarrow & \bfx_{sp}+\Delta tv_{x,p}\bfe_{x}~,\\
\bfv_{p} & \rightarrow & \bfv_{p}+\frac{q_{i}}{m_{i}}v_{x,p}\bfe_{x}\times\int_{0}^{\Delta t}dt'\bfB_{0}\left(\bfx_{p}+v_{x,p}t'\bfe_{x},t_{p}\right)~,\\
W_{p} & \rightarrow & W_{p}-\Delta t\frac{q_{i}}{m_{i}}\int_{0}^{\Delta t}dt'\frac{\partial\bfA_{0}\left(\bfx_{p}+v_{x,p}t'\bfe_{x},t_{p}\right)}{\partial t_{p}}~,\\
t_{p} & \rightarrow & t_{p}~.
\end{array}\right.\\
\Theta_{y}\left(\Delta t\right) & : & \left\{ \begin{array}{ccl}
\bfx_{p} & \rightarrow & \bfx_{sp}+\Delta tv_{y,p}\bfe_{y}~,\\
\bfv_{p} & \rightarrow & \bfv_{p}+\frac{q_{i}}{m_{i}}v_{y,p}\bfe_{y}\times\int_{0}^{\Delta t}dt'\bfB_{0}\left(\bfx_{p}+v_{y,p}t'\bfe_{y},t_{p}\right)~,\\
W_{p} & \rightarrow & W_{p}-\Delta t\frac{q_{i}}{m_{i}}\int_{0}^{\Delta t}dt'\frac{\partial\bfA_{0}\left(\bfx_{p}+v_{y,p}t'\bfe_{y},t_{p}\right)}{\partial t_{p}}~,\\
t_{p} & \rightarrow & t_{p}~.
\end{array}\right.\\
\Theta_{z}\left(\Delta t\right) & : & \left\{ \begin{array}{ccl}
\bfx_{p} & \rightarrow & \bfx_{sp}+\Delta tv_{z,p}\bfe_{z}~,\\
\bfv_{p} & \rightarrow & \bfv_{p}+\frac{q_{i}}{m_{i}}v_{z,p}\bfe_{z}\times\int_{0}^{\Delta t}dt'\bfB_{0}\left(\bfx_{p}+v_{z,p}t'\bfe_{z},t_{p}\right)~,\\
W_{p} & \rightarrow & W_{p}-\Delta t\frac{q_{i}}{m_{i}}\int_{0}^{\Delta t}dt'\frac{\partial\bfA_{0}\left(\bfx_{p}+v_{z,p}t'\bfe_{z},t_{p}\right)}{\partial t_{p}}~,\\
t_{p} & \rightarrow & t_{p}~.
\end{array}\right.
\end{eqnarray}
Using these exact solutions of the subsystems, we can compose symplectic
iteration schemes of the entire system. For instance, a 1st-order
symplectic scheme is 
\begin{eqnarray}
\Theta_{1}\left(\Delta t\right)=\Theta_{x}\left(\Delta t\right)\Theta_{y}\left(\Delta t\right)\Theta_{z}\left(\Delta t\right)\Theta_{V}\left(\Delta t\right)\Theta_{W}\left(\Delta t\right)~,\label{EqnFSTHS}
\end{eqnarray}
and a symmetric 2nd-order symplectic scheme can be built as 
\begin{eqnarray}
\Theta_{2}\left(\Delta t\right) & = & \Theta_{W}\left(\Delta t/2\right)\Theta_{x}\left(\Delta t/2\right)\Theta_{y}\left(\Delta t/2\right)\Theta_{z}\left(\Delta t/2\right)\Theta_{V}\left(\Delta t\right)\nonumber \\
 &  & \Theta_{z}\left(\Delta t/2\right)\Theta_{y}\left(\Delta t/2\right)\Theta_{x}\left(\Delta t/2\right)\Theta_{W}\left(\Delta t/2\right)~.\label{EqnHAMS2}
\end{eqnarray}
A $2(l+1)$-th order symplectic scheme can be constructed from a $2l$-th
order symplectic scheme as \cite{yoshida1990construction} 
\begin{eqnarray}
\Theta_{2(l+1)}(\Delta t) & = & \Theta_{2l}(\alpha_{l}\Delta t)\Theta_{2l}(\beta_{l}\Delta t)\Theta_{2l}(\alpha_{l}\Delta t)~,\\
\alpha_{l} & = & 1/(2-2^{1/(2l+1)})~,\\
\beta_{l} & = & 1-2\alpha_{l}~.
\end{eqnarray}

We should point out that there exist a discrete variational approach
that generates the same explicit schemes as $\Theta_{1}$, $\Theta_{2}$
and $\Theta_{2l}$. See Refs. \cite{xiao2018structure,Xiao2018} for
details.

\section{Simulations of ion Bernstein waves and drift wave instabilities\label{sec:4}}

We have implemented the 2nd-order explicit structure-preserving geometric
PIC algorithm given by \EQ{EqnHAMS2} in the \textsl{SymPIC} code
to simulate low-frequency electrostatic perturbations with fully kinetic
ions and adiabatic electrons. The Whitney interpolating maps are chosen
to be the same as those in Ref. \cite{xiao2016explicit}. As a benchmark
and test, the algorithm is applied to study the ion Bernstein waves.
It is then used to simulate the drift wave instability and ion temperature
gradient turbulence in a 2D slab geometry.

\subsection{Dispersion relation of ion Bernstein waves}

To simulate the IBWs in a homogeneous magnetized plasma, the follow
system parameters are chosen. External magnetic field is in the $z$-direction,
$\bfB_{0}=B_{0,z}\bfe_{z}$ with $B_{0,z}=2.5$T. Plasma density $n_{i,0}=1\EXP{18}\mathrm{m}^{-3}$,
and the thermal velocity of ions $v_{T,i,0}=7.23\EXP{-4}\mathrm{c}$,
where $\mathrm{c}$ is the speed of light in the vacuum. The mass
and charge of ions are $m_{i}=3.342\EXP{-27}$kg and $q_{i}=1.6\EXP{-19}$C,
respectively. The simulation domain is a $256\times1\times1$ grid
and periodic boundaries are imposed for all 3 directions. On average
there are 256 simulation particles (sampling points) per grid cell.
The grid sizes are $\Delta x=4\EXP{-4}\mathrm{m}=0.2208\rho_{T,i}$
and $\Delta t=120\Delta x/\mathrm{c}=0.0192/\omega_{c,i}~.$ Here
the time-step $\Delta t$ is relatively small compared with the cyclone
period $2\pi/\omega_{c,i}$, because it needs to satisfy the Courant
condition for stability. Initially the perturbed electromagnetic fields
is set to zero, and electrostatic waves are generated from noise.
The total number of time-steps is 8192.

\begin{figure}
\begin{centering}
\includegraphics[width=0.6\linewidth]{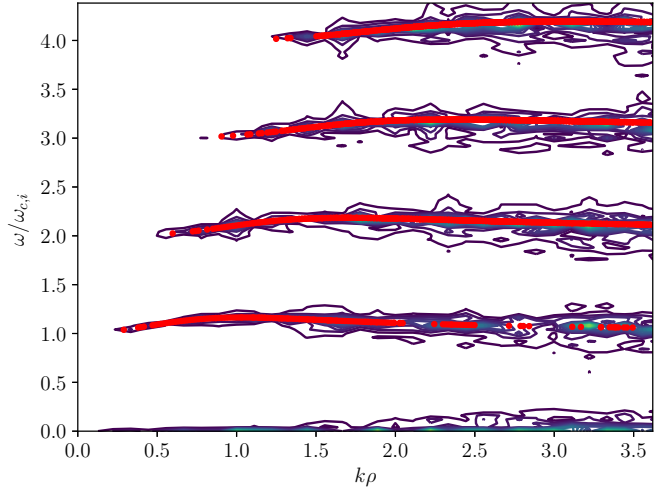} 
\par\end{centering}
\caption{Dispersion relation of ion Bernstein waves in a hot magnetized plasma
simulated by structure-preserving geometric PIC algorithm. Red dots
are analytical dispersion relation.}
\label{FigIBWDSP} 
\end{figure}

Theoretically the dispersion relation of the electrostatic IBWs in
the $x$-direction is \cite{miecnikowski2018nonlinear,fried1968plasma}
\begin{eqnarray}
\epsilon\left(\omega,k\right) & = & 1+\frac{\theta}{\rho_{T,i}}\sum_{n=-\infty}^{\infty}nI_{n}\left(b\right)\exp\left(-b\right)\frac{v_{T,i,0}}{\omega+n\omega_{c,i}}~,
\end{eqnarray}
where 
\begin{eqnarray*}
b & = & k^{2}\rho_{T,i}^{2}~,\quad\theta=T_{e}/T_{i}=1~,\quad T_{i}=m_{i}v_{T,i,0}^{2}~,\\
\omega_{c,i} & = & q_{i}B_{0,z}/m_{i}~,\quad\rho_{T,i}=m_{i}v_{T,i,0}/(q_{i}B_{0,z})~.
\end{eqnarray*}
The spectra of the electric field in the $x$-direction is plotted
in \FIG{FigIBWDSP}, which clearly shows that the simulated dispersion
relation matches the theoretical result very well. 

\begin{figure}[h]
\begin{centering}
\includegraphics[width=0.6\linewidth]{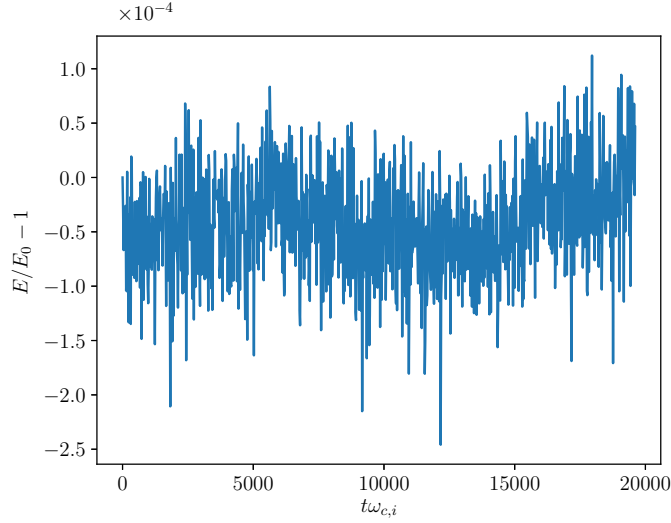} 
\par\end{centering}
\caption{The energy error of structure-preserving geometric PIC algorithm is
bounded by a small number for all simulation time-steps.}
\label{FigIBWENE}
\end{figure}

To test the energy conservation property, we performed a long-term
simulation. The total number of time-steps is $1\times10^{6}$. The
simulation domain is a $32\times32\times32$ grid mesh, and the averaged
number of simulation particles per cell is 16. During the simulation
the total energy is recorded, and the result is shown in \FIG{FigIBWENE}.
It is evident that the error of total energy is bounded by a small
number for all simulation time-steps.

\label{SecIBW}

\subsection{Ion temperature gradient instability and turbulence in a slab geometry}

\begin{figure}[h]
\begin{centering}
\includegraphics[width=0.6\linewidth]{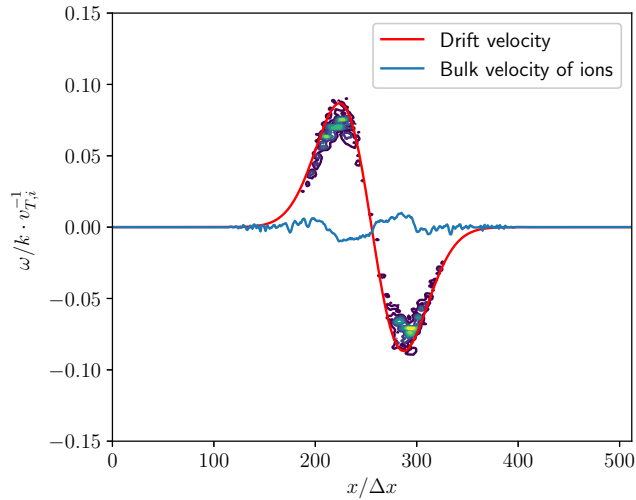}
\par\end{centering}
\caption{Phase velocity in the $y$-direction as a function of $x$ at $k_{y}=40\pi/\left(512\Delta x\right)$
and $t=766/\omega_{c,i}$. Solid lines are the theoretical value of
$v_{d,y}\left(x\right)$ (red) given by Eq.\,(\ref{eq:vd}) and the
bulk velocity of the ions in the $y$-direction (blue).}
\label{FigDWK20}
\end{figure}

In certain parameter regimes, the ion temperature gradient in a magnetized
plasma can excite the drift wave instability, which often nonlinearly
evolves into a turbulent stage to produce anomalous transport of energy
and particles \cite{Krall1965,Coppi1967,Tang1978,romanelli1989ion,Hammett1990,cowley1991considerations,horton1999drift,dorland1993gyrofluid,parker1999large,Rogers2000,Dimits2007,Ku2009,Merz2010,sturdevant2017low,miecnikowski2018nonlinear,hu2018fully}.
We demonstrate the simulations of the ITG instability and turbulence
by the structure-preserving geometric PIC algorithm in a 2D slab geometry.
System parameters are similar to those in \SEC{SecIBW}, except
that the temperatures for both ions and electrons are now functions
of the $x$-coordinate,
\begin{eqnarray}
T_{e}\left(x\right)=T_{i}\left(x\right) & = & m_{i}v_{T,i,0}^{2}\exp\left(-\frac{\left(x-x_{m}\right)^{2}}{\sigma^{2}}\right)~,\\
x_{m} & = & 256\Delta x~,\,\,\,\,\sigma=44.72\Delta x~.
\end{eqnarray}
The simulation domain is a $N_{x}\times N_{y}\times N_{z}=512\times512\times1$
grid, and the total number of time-steps is $1.2\times10^{6}$. To
balance the pressure gradient for equilibrium, we use an external
electric field, which is set to 
\begin{eqnarray}
\bfE_{0}\left(x,y,z\right) & = & \frac{\partial T_{i}\left(x\right)}{\partial x}\frac{1}{q_{i}}~.
\end{eqnarray}
It can be checked that the local Maxwell distribution function
\begin{equation}
f_{0}\left(\bfx,\bfv\right)=\frac{n_{0}}{\left(2\pi T_{i}\left(x\right)/m_{i}\right)^{3/2}}\exp\left(-\frac{|\bfv|^{2}}{T_{i}\left(x\right)/m_{i}}\right)~,\label{eq:f0}
\end{equation}
is the steady state solution of the $0$th, 1st, and 2nd-order moment
equations of the Vlasov equation, i.e.,
\begin{eqnarray*}
\int\rmd v^{3}\left(\bfv\cdot\nabla f_{0}+\frac{q_{i}}{m_{i}}\left(\bfE_{0}+\bfv\times\bfB_{0}\right)\frac{\partial}{\partial\bfv}f_{0}\right) & = & 0~,\\
\int\rmd v^{3}\left(\bfv\bfv\cdot\nabla f_{0}+\bfv\frac{q_{i}}{m_{i}}\left(\bfE_{0}+\bfv\times\bfB_{0}\right)\frac{\partial}{\partial\bfv}f_{0}\right) & = & 0~,\\
\int\rmd v^{3}\left(\bfv\bfv\bfv\cdot\nabla f_{0}+\bfv\bfv\frac{q_{i}}{m_{i}}\left(\bfE_{0}+\bfv\times\bfB_{0}\right)\frac{\partial}{\partial\bfv}f_{0}\right) & = & 0~.
\end{eqnarray*}
To obtain a more precise kinetic equilibrium, we first use the $f_{0}$
specified by Eq.\,(\ref{eq:f0}) to perform a 1-D simulation, i.e.,
$N_{y}=1$. After $10^{6}$ time-steps when the ion distribution function
reaches a steady state, we take this numerically calculated distribution
function as the equilibrium distribution function for the 2D simulation
in the slab geometry. For the system parameters selected in this example,
the ion temperature gradient excites unstable drift modes. According
to the theory of drift wave, the phase velocity in the $y$-direction
of modes is approximately 
\begin{eqnarray}
v_{d,y}\left(x\right) & = & \frac{\partial T_{i}\left(x\right)}{\partial x}\frac{1}{m_{i}\omega_{c,i}}~.\label{eq:vd}
\end{eqnarray}
Plotted in \FIG{FigDWK20} are the phase velocity in the $y$-direction
calculated from the electric field perturbations observed in the simulation
and the theoretical drift velocity $v_{d,y}\left(x\right)$ as a function
of $x$ given by Eq.\,(\ref{eq:vd}). It is clear that the simulation
agrees with the theoretical predication very well. We also plotted
in \FIG{FigDWK20} the averaged bulk velocity of the ions in the
$y$-direction, which by comparison is smaller. This indicates that
the space-time structure observed in the simulation is produced by
the drift wave, instead of the bulk flow of the ions. 

\begin{figure}[h]
\begin{centering}
\includegraphics[width=1\linewidth]{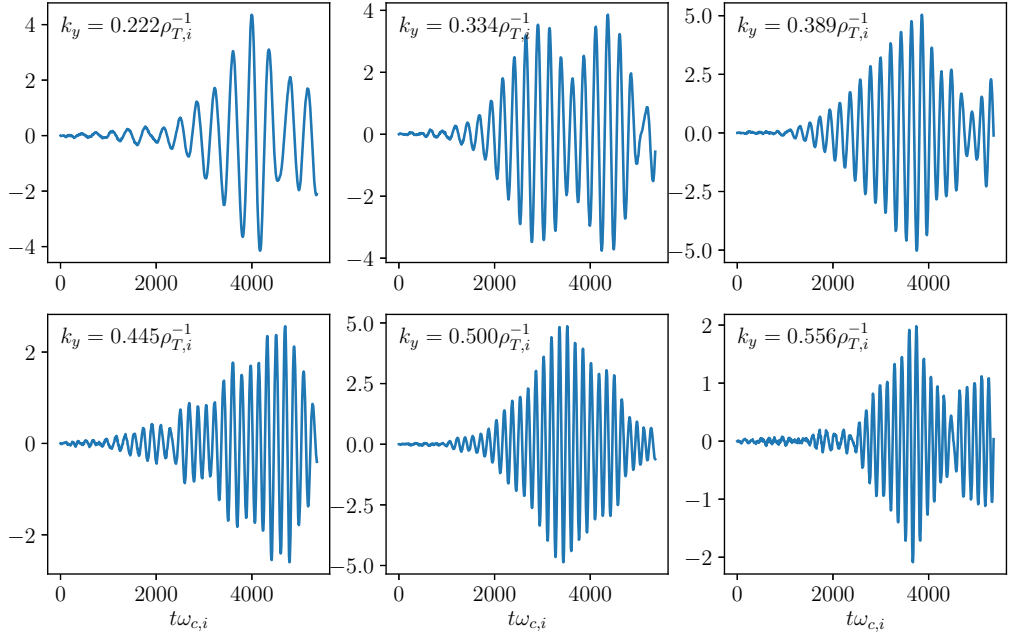} 
\par\end{centering}
\caption{Time history of the amplitude of the unstable modes at $x=224\Delta x$
for different values of $k_{y}.$}
 \label{FigITGPLOT} 
\end{figure}

\begin{figure}
\begin{centering}
\includegraphics[width=0.6\linewidth]{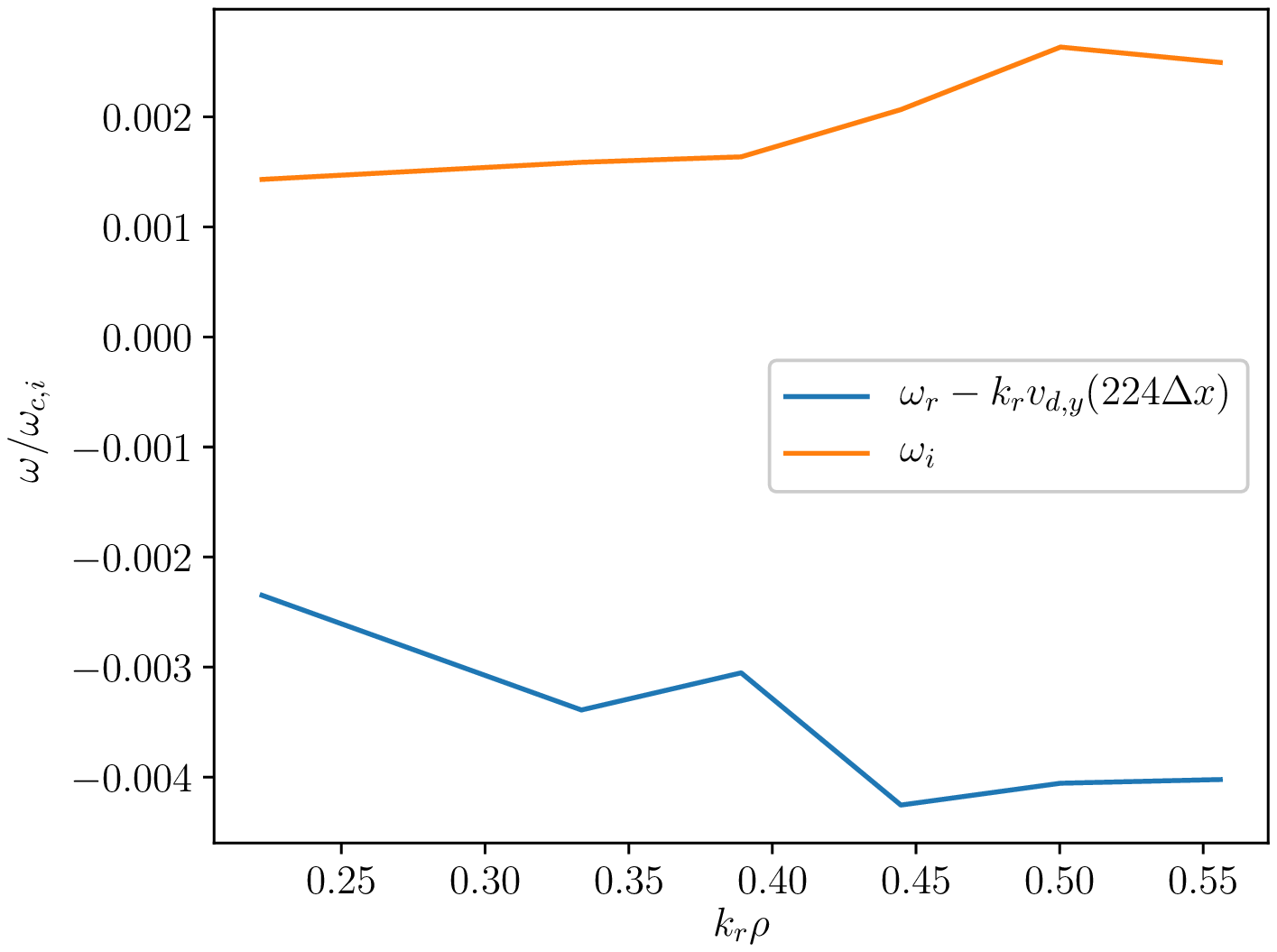} 
\par\end{centering}
\caption{Dispersion relation of unstable modes at $x=224\Delta x$ calculated
from simulation data.}
\label{FigGAMMA} 
\end{figure}

To illustrate the instability, the time history of the amplitude of
density perturbation with different $k_{y}$ at $x=224\Delta x$ are
plotted in \FIG{FigITGPLOT}. We observe that all modes displayed
grow initially, and saturate after $t>3000/\omega_{c,i}$. From the
simulation data, the dispersion relation of the instability at $x=224\Delta x$
can be calculated. It is plotted in \FIG{FigGAMMA}.

\begin{figure}
\begin{centering}
\includegraphics[width=1\linewidth]{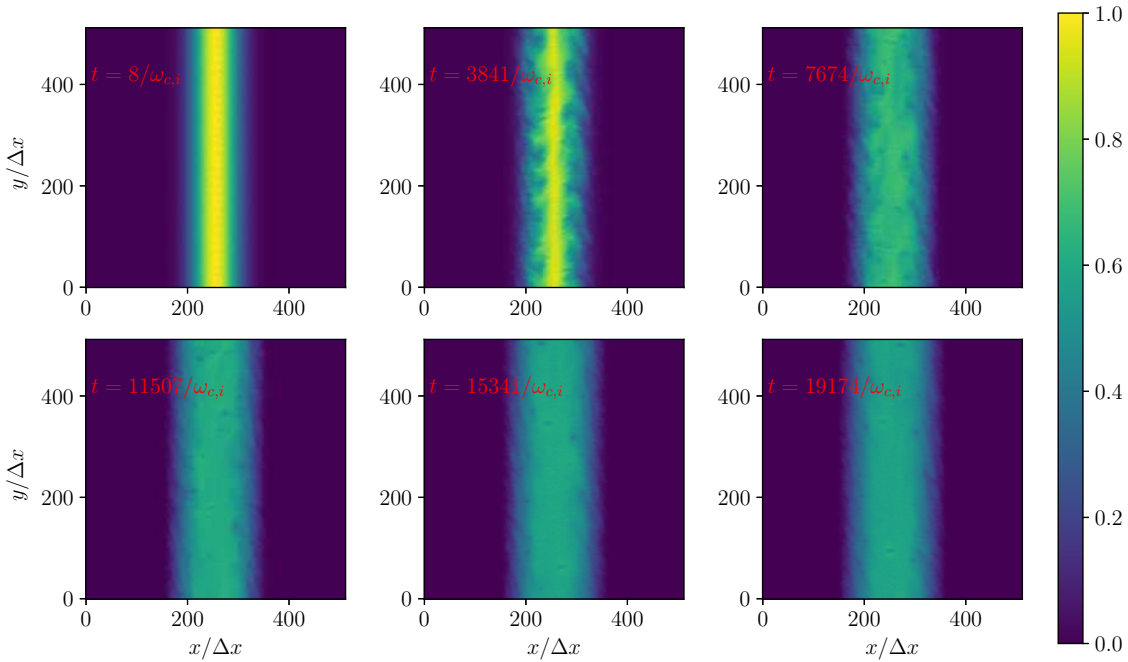} 
\par\end{centering}
\caption{The kinetic energy diffuses as the instability grows, saturates, and
becomes turbulent.}
\label{FigITDT} 
\end{figure}

\begin{figure}
\begin{centering}
\includegraphics[width=1\linewidth]{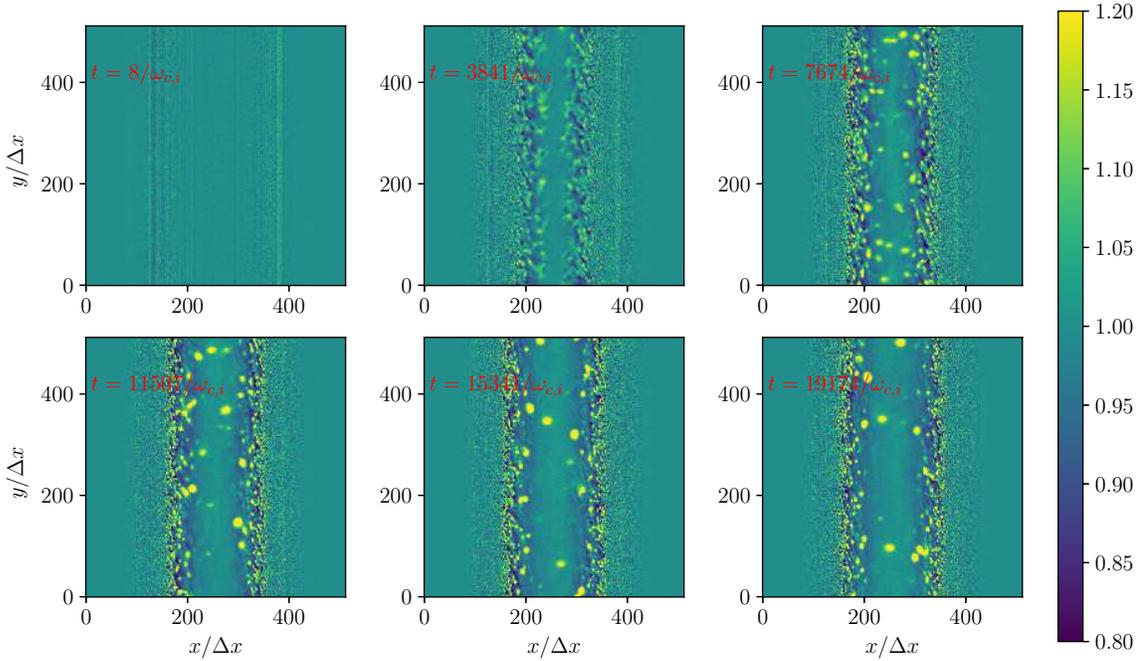} 
\par\end{centering}
\caption{Density perturbation at different times. Density blobs generated by
the rupture of unstable modes are the prominent structures of the
fully developed ITG turbulence.}
\label{FigITDN} 
\end{figure}

After a sufficient long time, the unstable modes nonlinearly evolve
into a turbulent state, as evident from the distribution of kinetic
energy density and number density of ions at different times. Figure
\ref{FigITDT} shows that the kinetic energy diffuses as the instability
grows, saturates, and becomes turbulent. Figure \ref{FigITDN} shows
that density blobs generated by the rupture of unstable modes are
the prominent structures of the fully developed ITG turbulence. The
details of the instability and turbulence, especially the formation
of density blobs, can be observed from the video of the density evolution
available at \url{http://staff.ustc.edu.cn/~xiaojy/ditg.html}.

\begin{figure}
\begin{centering}
\includegraphics[width=1\linewidth]{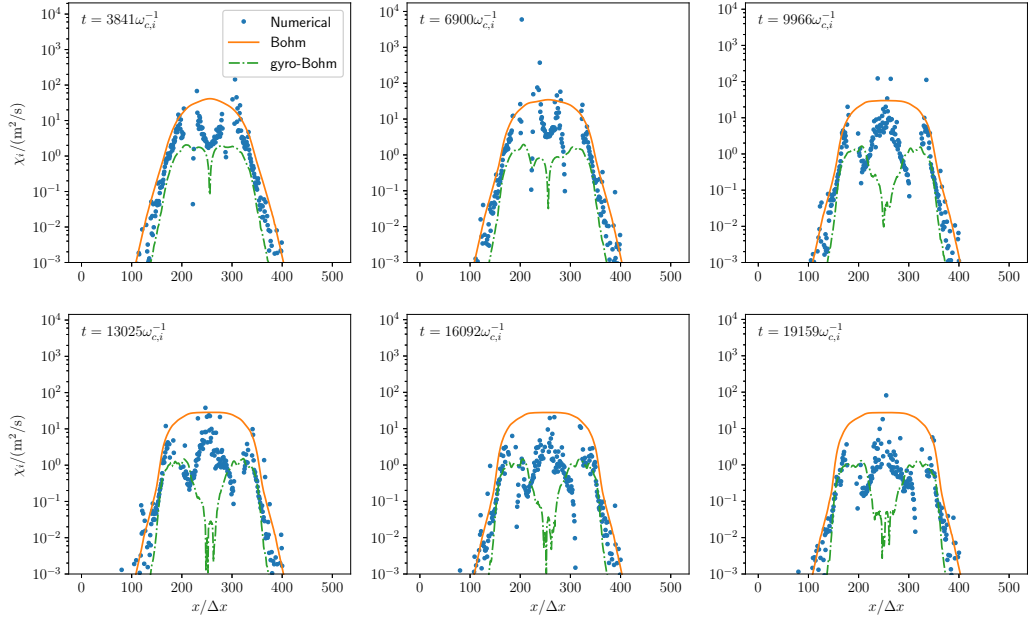} 
\par\end{centering}
\caption{Energy diffusion coefficient in the $x$-direction at different times.
The diffusion coefficient is closer to the gyro-Bohm scaling when
the ITG turbulence is fully developed.}
\label{FigDCT} 
\end{figure}

When turbulence develops, the energy or particle diffusion of the
plasma across the magnetic field is conjectured empirically to follow
the scaling of Bohm diffusion or the gyro-Bohm diffusion \cite{taylor1961diffusion,drummond1962anomalous,waltz1990magnetic,petty1995nondimensional}.
The corresponding diffusion coefficients are
\begin{eqnarray}
\chi_{B} & = & \frac{k_{B}T_{i}}{16q_{i}B}~,\\
\chi_{gB} & = & \rho^{*}\chi_{B}~,
\end{eqnarray}
where $\rho^{*}=\rho_{T,i}/L$ is the gyro-radius of ions measured
in $L$, the characteristic length of the plasma. Assuming that the
diffusion coefficient varies slowly with $x$ and the energy density
$E_{k}$ diffuses according to 
\begin{eqnarray}
\dot{E_{k}} & = & \chi\frac{\rmd^{2}E_{k}}{\rmd x^{2}}~,
\end{eqnarray}
we can calculate the numerical diffusion coefficient of ions $\chi$
in the $x$-direction. The results are plotted in \FIG{FigDCT},
where the local plasma characteristic length $L$ is estimated using
\begin{eqnarray}
L=\frac{E_{k}}{\rmd E_{k}/\rmd x}~.
\end{eqnarray}
We observe that at $t=3841/\omega_{c,i}$ the diffusion coefficient
$\chi$ is between the Bohm scaling $\chi_{B}$ and gyro-Bohm scaling
$\chi_{gB}$. Afterwards, $\chi$ decreases. When the ITG turbulence
is fully developed at $t=19159/\omega_{c,i}$, the diffusion coefficient
$\chi$ is closer to the gyro-Bohm scaling $\chi_{gB}$.

\begin{figure}
\begin{centering}
\includegraphics[width=0.6\linewidth]{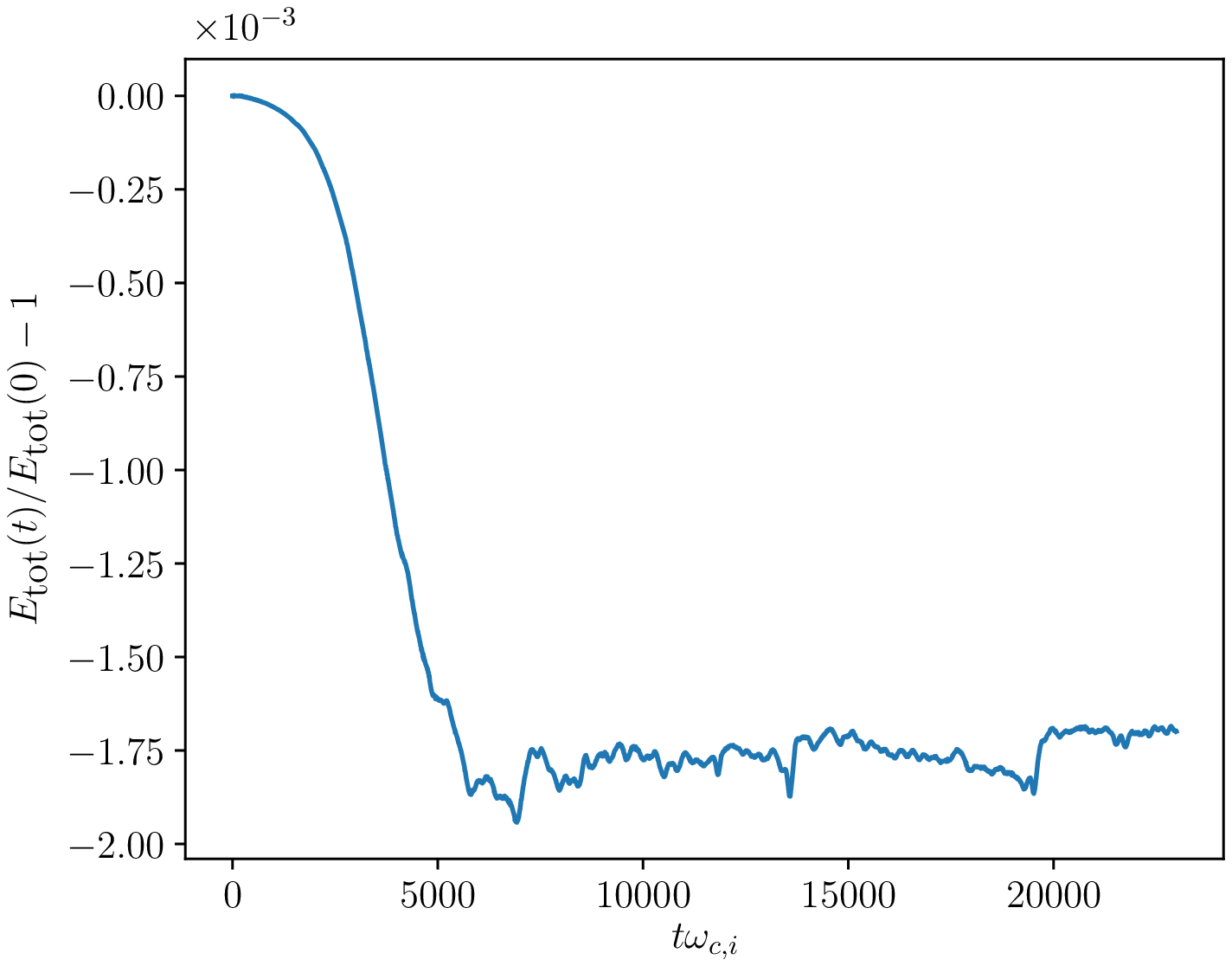}
\par\end{centering}
\caption{Time history of the energy error of ITG turbulence simulation using
the structure-preserving geometric PIC algorithm. The error is globally
bounded by a small number for all simulation time-steps.}
\label{FigENEA} 
\end{figure}

We emphasize again that simulating long-term dynamical and transport
behavior of magnetized plasmas using fully kinetic particles demands
a large number of simulation time-steps, $1.2\times10^{6}$ in this
case. A structure-preserving geometric algorithm with long-term accuracy
and fidelity is desirable for this purpose. To verify the long-term
conservative properties of the simulation, we plotted the energy error
in \FIG{FigENEA}. Clearly, the error is globally bounded by a small
number for all simulation time-steps.

\section{Discussion and Conclusion}

In conclusion, we have customarily designed a field theory for low
frequency electrostatic perturbations with fully kinetic ions and
adiabatic electrons, and geometrically discretized the field theory
to build a structure-preserving geometric PIC algorithm for simulating
drift wave instabilities and turbulence in magnetized plasmas. The
geometric discretization of the field theory is accomplished using
DEC, high-order Whitney interpolation forms, and the non-canonical
Hamiltonian splitting method. It preserves the non-canonical symplectic
structure of the particle-field system, as well as the gauge symmetry.
And as a result, the PIC algorithm is automatically charge-conserving
and possesses long-term conservation properties that are indispensable
for simulating the dynamics of fully kinetic particles. We have successfully
implemented the algorithm in the \textsl{SymPIC} code. The algorithm
was tested and benchmarked using the examples of ion Bernstein waves
and drift waves, and applied to study the ion temperature gradient
instability and turbulence in a 2D slab geometry. Simulation results
show that at the early stage of the ITG turbulence, the energy diffusion
is between the Bohm scaling and gyro-Bohm scaling. At later time,
the observed diffusion is closer to the gyro-Bohm scaling, and density
blobs generated by the rupture of unstable modes are the prominent
structures of the fully developed ITG turbulence.
\begin{acknowledgments}
This research was supported by the National Key Research and Development
Program (2016YFA0400600, 2016YFA0400601 and 2016YFA0400602), the National
Natural Science Foundation of China (NSFC-11775219 and NSFC-11575186),
China Postdoctoral Science Foundation (2017LH002), the Fundamental
Research Funds for the Central Universities (WK2030040096) and the
U.S. Department of Energy (DE-AC02-09CH11466).
\end{acknowledgments}

\bibliographystyle{apsrev4-1}
\bibliography{vsitg}

\end{document}